\documentstyle[aps,prb]{revtex}
\begin{document}

\title{On the theory of polarized Fermi liquid} 
\author{V.P.Mineev}
\address{Commissariat a l'Energie Atomique, DSM, Departement de
Recherche Fondamentale sur la Matiere Condensee, SPSMS, 38054
Grenoble, France } 
\date{ Submitted 1 May 2004}
\maketitle
\begin{abstract}
The transport equation for transverse vibrations of magnetization in spin
polarized Fermi liquid is derived from integral equation for the
vertex function.  The dispersion law for the transverse spin waves is
established.  The existance of zero-temperature spin-waves attenuation
is confirmed.  The problem of similar derivation in ferromagnetic 
"Fermi liquid" is discussed.
    
\end{abstract}

\bigskip

\section{Introduction}

It is well known that the phenomenological Landau Fermi liquid theory has
well established foundations based on microscopic theory.  Namely, the
transport equation for the vibrations of a Fermi liquid was derived
from an integral equation for vertex function and general relation
between the vertex function and quasiparticle interaction function was
found \cite{1}.  The goal of the present article is to make similar
derivation for a spin-polarized Fermi liquid.  The principal interest
of this consists in establishing of a general basis for so called
zero-temperature spin-wave relaxation observed in spin-echo
experiments \cite{2}.  The transverse spin-wave damping in a Fermi
liquid with finite polarization  was calculated in many theoretical
investigations \cite{3,4,5,6,7,8}.  On the other hand its existance
at $T=0$ has been contested in the paper \cite{9} (see also \cite{8} for
discussion).

Unlike to the previous treatments based either on microscopic
approaches developed for a diluted Fermi gas \cite{3,4,5,6} or on
phenomenological description of polarized Fermi liquid \cite{7,8} we
shall consider here a system of fermions at $T=0$, with arbitrary
short range interaction forces.  The presence of polarization means
that the particle distribution functions for spin-up and spin-down
particles have different Fermi momenta $p_{+}¥$ and $p_{-}¥$.  The
Green functions near ${\bf p}=p_{\pm}¥$ and $\varepsilon=\mu_{\pm}¥$
have the form
\begin{equation}
G_{\pm}¥({\bf p}, \varepsilon)=\frac{a}{\varepsilon -
\mu_{\pm}¥-v_{F}¥(p-p_{\pm}¥) +ib(p-p_{\pm}¥)|p-p_{\pm}¥|}.
\label{e1}
\end{equation}

We shall assume a weak polarization
$v_{F}¥(p_{+}¥-p_{-}¥)\ll\varepsilon_{F}¥$ and also that both the Fermi
distributions are characterized by the same Landau Fermi liquid
parameters.  Unlike \cite{10} we introduce here the general form of
imaginary part of self-energy \cite{11} which is quadratic function of the
difference $(p-p_{\pm}¥)$ and changes its sign at $p=p_{\pm}¥$
correspondingly.  The assumption of small polarization in particular
means that $G_{+}¥$ is given by expression (\ref{e1}) not only near
its own Fermi surface ${\bf p}=p_{+}¥$ and $\varepsilon=\mu_{+}¥$ but
in whole intervals $p_{-}¥<p<p_{+}¥$ and
$\mu_{-}¥<\varepsilon<\mu_{+}¥$ and also near the "alien" Fermi
surface ${\bf p}=p_{-}¥$ and $\varepsilon=\mu_{-}¥$.  The same is
truth for $G_{-}¥$.

In general the polarization is
nonequilibrium, hence $\mu_{+}¥-\mu_{-}¥=\Omega-\omega_{L}¥$, where
$\omega_{L}¥=\gamma H_{0}¥$ is the Larmor frequency corresponding to
the external field $H_{0}¥$ and $\Omega$ is the Larmor frequency
corresponding to the effective field which would produce the existing
polarization , $v_{F}¥(p_{+}¥-p_{-}¥)= \Omega/(1+F_{0}¥^{a}¥)$.
\cite{12}

Following Landau let us write equation for the vortex function for the
scattering of two particles with opposite spin direction and a small
transfer of 4-momentum $K=({\bf k},\omega)$
\begin{equation}
\Gamma(P_{1}¥,P_{2}¥,K)=\Gamma_{1}¥(P_{1}¥,P_{2}¥)-\frac{i}{(2\pi)^{4}¥}
\int \Gamma_{1}¥(P_{1}¥,Q)G_{+}¥(Q)G_{-}¥(Q+K)\Gamma(Q,P_{2}¥,K)d^{4}¥Q
\label{e2}
\end{equation}
If $K$ is small and polarization is also small, the poles of two Green
functions are close each other.
Let us assume that
all other quantities in the integrand are slowly varying with respect
to $Q$: their energy and momentum scales of variation are larger than $\max
\{\Omega, \omega\}$ and $\max\{\Omega/v_{F}¥,k\}$ correspondingly.
Then one can perform the integration in (\ref{e2}) at fixed values of 
$Q=p_{0}¥=(p_{+}¥+p_{-}¥)/2,\mu=(\mu_{+}¥+\mu_{-}¥)/2$ in the arguments
of $\Gamma$ and $\Gamma_{1}¥$ functions.  Another words, one can
substitute in (\ref{e2})
\begin{eqnarray}
&G_{+}¥&(Q)G_{-}¥(Q+K)=G_{+}¥({\bf q},\varepsilon)
G_{-}¥({\bf q}+{\bf k},\varepsilon+\omega)\\ \nonumber
&=&\frac{2\pi i a^{2}¥}{v_{F}¥}\delta(\varepsilon-\mu)\delta(|{\bf q}|-p_{0}¥)
\frac{\frac{\Omega}{1+F_{0}¥^{a}¥}+{\bf k}{\bf v}_{F}¥}{\omega-\omega_{L}¥
+\frac{\Omega F_{0}¥^{a}¥}{1+F_{0}¥^{a}¥} -
{\bf k}{\bf v}_{F}¥+ 
\frac{ib{\bf k}{\bf v}_{F}¥\Omega}{v_{F}¥^{2}¥(1+F_{0}¥^{a}¥)}}+
\Phi_{\mbox{reg}}¥.
\label{e3}
\end{eqnarray}
We have neglected in denominator by the terms of the second order on
polarization.  Considering now  two
limits $\Gamma^{\omega}¥= \Gamma(\omega\to\omega_{L}¥, |{\bf
k}|/(\omega-\omega_{L}¥)\to 0, \Omega= 0)$ and 
$\Gamma^{\bf k}¥=\Gamma(|{\bf k}|\to 0,(\omega-\omega_{L}¥)/|{\bf k}|\to 0,
\Omega = 0)$ we come, according to well known procedure \cite{1}, to kinetic equation
\begin{equation}
\left (\omega-\omega_{L}¥+\frac{\Omega
F_{0}¥^{a}¥}{1+F_{0}¥^{a}¥}- {\bf k}{\bf n}v_{F}¥ + \frac{ib{\bf k}{\bf n}
v_{F}¥\Omega}{v_{F}¥^{2}¥(1+F_{0}¥^{a}¥)}\right )\nu({\bf n})=
\left (\frac{\Omega}{1+F_{0}¥^{a}¥}+{\bf k}{\bf n}v_{F}¥\right) 
\int\frac{d{\bf n'}¥}{4\pi} F_{{\bf n}{\bf n'}}\nu({\bf n'})¥.
\label{e4}
\end{equation}
We limit ourselves by the first two harmonics in the Landau interaction
function $F_{{\bf n}{\bf n'}}=F_{0}¥^{a}¥+({\bf n}{\bf n'})F_{1}¥^{a}¥$.
Introducing the expansion of the distribution function $\nu({\bf n})¥$
over spherical harmonics of direction ${\bf n}={\bf v}_{F}¥/v_{F}¥$
one can find from (\ref{e4}) that the ratio of amplitudes of the
successive harmonics with $l\ge 1$ is of the order of
$kv_{F}¥/\Omega$.  Hence if it is assumed this ratio as a small parameter
one can work with distribution function taken in the form \cite{13}
$\nu({\bf n})=\nu_{0}¥+({\bf n}\hat{\bf k})\nu_{1}¥$.
The functions $\nu_{0}¥$ and $\nu_{1}¥$ obey the following system of
linear equations:
\begin{equation}
( \omega-\omega_{L}¥ )\nu_{0}¥-\frac{kv_{F}¥}{3}
\left (1+\frac{F_{1}¥^{a}¥}{3}
-\frac{ib{\bf k}{\bf v}_{F}¥\Omega}{v_{F}¥^{2}¥(1+F_{0}¥^{a}¥)}  \right )
\nu_{1}¥=0,
\label{e5}
\end{equation}
\begin{equation}
-kv_{F}¥\left (1+F_{0}¥^{a}¥
-\frac{ib{\bf k}{\bf v}_{F}¥\Omega}{v_{F}¥^{2}¥(1+F_{0}¥^{a}¥)} \right)\nu_{0}¥+
\left (\omega-\omega_{L}¥+\frac{\Omega
(F_{0}¥^{a}¥-\frac{F_{1}¥^{a}¥}{3})}{1+F_{0}¥^{a}¥}\right )\nu_{1}¥=0.
\label{e6}
\end{equation}
The equality to zero of the determinant of this system gives the spin
waves dispersion law.  At long enough wave lengths when the dispersive
part of $\omega(k)$ dependence is much less than $\omega_{L}¥$ we have
\begin{equation}
\omega=\omega_{L}¥+ (D^{\prime\prime}¥-iD^{\prime}¥)k^{2}¥,
\label{e7}
\end{equation}
where 
\begin{equation} 
D^{\prime\prime}¥=\frac{v_{F}¥^{2}¥(1+F_{0}¥^{a}¥)(1+F_{1}¥^{a}¥/3)}
{3 \kappa\gamma H}
\label{e8}
\end{equation}
is the reactive part of diffusion coefficient,
\begin{equation}     
D^{\prime}¥=\frac{b\left (2+F_{0}¥^{a}¥+\frac{F_{1}¥^{a}¥}{3}\right )}
{3\kappa}
\label{e9}
\end{equation}
is dissipative part of diffusion coefficient,
$\kappa=F_{0}¥^{a}¥-F_{1}¥^{a}¥/3$ and
$H=\Omega/\gamma(1+F_{0}¥^{a}¥)$ is  effective "internal" field
corresponding to effective "external" field $\Omega/\gamma$ producing
the existing polarization.

The expressions for $D^{\prime}¥$ and $D^{\prime\prime}¥$ found here reproduce
the corresponding expressions have been obtained from phenomenological
Landau-Silin kinetic equation with two-particle collision integral
\cite{8} at arbitrary relation between polarization and temperature if
we put in the latters $T=0$ .  In particular $D^{\prime}¥$ proves to
be polarization independent whereas $D^{\prime\prime}¥$ is
inversely proportional to polarization.

Thus general microscopic derivation  confirms the statement about the
existance of zero-temperature spin waves attenuation in polarized
Fermi liquid.  The value of the dissipative part of spin diffusion 
$D^{\prime}¥$ is determined by the amplitude "b" of the imaginary
self-energy function.  Hence it originates of collisions between
quasiparticles. 

The derivation for the transverse spin waves proposed here is relevant to
spin-polarized Fermi-liquid.  It is appropriate to look on the same
problem for "Fermi liquid" with spontaneous magnetization.  This
subject has been discussed first phenomenologically by A.A.Abrikosov
and I.E.Dzyaloshinskii \cite{14} and after microsopically by
P.S.Kondratenko \cite{15}.  They did not include in the theory a
finite scattering rate between quasiparticles and as result they have
obtained the dissipationless transvese spin wave dispersion law
as it seemed to be in isotropic ferromagnet.  The derivation
\cite{14} has been critisized by C.Herring \cite{16} who pointed out
on the existance of the finite scattering rate and inapplicability of
naive Fermi-liquid approach to itinerant ferromagnet (see discussion
in \cite{8}).  Later I.E.Dzyaloshinskii and P.S.Kondratenko \cite{17}
have rederived the spin-wave dispersion law in ferromagnets.  Making
use as the starting point the Landau equation for the vertex function
for the scattering of two particles with opposite spin direction and a
small transfer of 4-momentum they have redefined the product of two
Green functions $G_{+}¥G_{-}¥$ in such a manner that its resonant part
was taken equal to zero at $\omega=0$ that is in fact equivalent to
interchange of the roles of $\Gamma^{\omega}¥$ and $\Gamma^{\bf k}¥$
functions.  Then taking into account the $1/k^{2}¥$ divergency of
transverse static susceptibility in the ferromagnetically ordered
state they have found the dissipationless transvese spin wave
dispersion law
\begin{equation}     
\omega=v_{F}¥(p_{+}¥-p_{-}¥) (ck)^{2}¥.    
\label{e10}
\end{equation}
Unlike to polarized Fermi liquid , where reactive part of
diffusion constant (\ref{e8}) is inversely proportional to
polarization, the constant in the ferromagnets spin-wave dispersion law is
proportional to polarization as should be according to
phenomenological Landau-Lifshits equation \cite{18}.  As in previous
papers \cite{14,15} the authors of \cite{17} did not introduce a
scattering rate in the momentum space between the Fermi surfaces for
the particles with opposite spins.  However, the literal reproduction of the
derivation proposed in \cite{17} with the Green functions (\ref{e1})
taking into account the finite quasiparticle scattering rate in whole
intervals $p_{-}¥<p<p_{+}¥$ and $\mu_{-}¥<\varepsilon<\mu_{+}¥$ leads,
nevertheless, to the same dissipationless transverse spin wave
dispersion law (\ref{e10}).  This result looks like as an inevitable
consequence of using in the process of derivation the singularity of
the transverse static susceptibility.  The latter of course is an
inherent property of a ferromagnet but, this property does not follow
from Fermi liquid theory as itself.  So in our opinion the
corresponding to Goldstone theorem dissipationless transverse spin
wave dispersion law in isotropic ferromagnet is certainly valid but
can not be derived on the basis of Fermi liquid theory only.

In conclusion, we note that making use the quantum-field theoretical approach
originally introduced by Landau, we have derived the dispersion law
for the transverse spin waves in a weakly polarized Fermi liquid at
$T=0$.  Along with the dissipationless part it contains also the
finite zero-temperature damping.  The polarization dependence both
dissipative and reactive part of diffusion constants corresponds to
dependences found earlier by means of kinetic equations with
two-particle collision integral.


\end{document}